\documentclass[10pt,aps,pre,twocolumn,showkeys,groupedaddress,floatfix,longbibliography]{revtex4-1}

\usepackage{graphicx}
\usepackage{amsmath}
\usepackage{amssymb}
\usepackage{algcompatible}
\usepackage{hyperref}
\begin{document}

\title[MC Simulation of Evaporation]{Monte Carlo Simulation of Evaporation Driven Self-Assembly in Suspension of Colloidal Rods}
\author{Nikolai I. Lebovka}
\email[Correspondence author: ]{lebovka@gmail.com}
\affiliation{F.D. Ovcharenko Institute of Biocolloidal Chemistry, NAS of Ukraine, Kiev, Ukraine}
\affiliation{Taras Shevchenko Kiev National University, Department of Physics, Kiev, Ukraine}
\author{Nikolai V. Vygornitskii}
\affiliation{F.D. Ovcharenko Institute of Biocolloidal Chemistry, NAS of Ukraine, Kiev, Ukraine}
\author{Volodymyr A. Gigiberiya}
\affiliation{F.D. Ovcharenko Institute of Biocolloidal Chemistry, NAS of Ukraine, Kiev, Ukraine}
\author{Yuri Yu. Tarasevich}
\email[Correspondence author: ]{tarasevich@asu.edu.ru}
\affiliation{Astrakhan State University, Astrakhan, Russia}
\date{\today}

\begin{abstract}
The vertical drying of a colloidal film containing rod-like particles was studied by means of kinetic Monte Carlo (MC) simulation.
The problem was approached using a two-dimensional square lattice and the rods were represented as linear $k$-mers (i.e., particles occupying $k$ adjacent sites). The initial state before drying was produced using a model of random sequential adsorption (RSA) with isotropic orientations of the $k$-mers (orientation of the $k$-mers along horizontal $x$ and vertical $y$ directions are equiprobable). In the RSA model, overlapping of the $k$-mers is forbidden. During the evaporation, an upper interface falls with a linear velocity of $u$ in the vertical  direction and the $k$-mers undergo translation Brownian motion.  The MC simulations were run at different initial concentrations, $p_i$, ($p_i \in [0, p_j]$, where $p_j$ is the jamming concentration), lengths of $k$-mers ($k \in [1, 12]$), and solvent evaporation rates, $u$. For completely dried films, the spatial distributions of $k$-mers and their electrical conductivities in both $x$ and $y$  directions were examined. Significant evaporation-driven self-assembly and orientation stratification of the $k$-mers oriented along the $x$ and $y$ directions were observed. The extent of stratification increased with increasing value of $k$. The anisotropy of the electrical conductivity of the film can be finely regulated by changes in the values of $p_i$, $k$ and $u$.
\end{abstract}

\keywords{computer simulation, drying, stratification, rods, electrical conductivity, anisotropy}


\maketitle

\section{Introduction\label{sec:intro}}

Vertical drying is an effective evaporation driven self-assembly (EDSA) technique for the organization of colloidal particles in thin films~\cite{Routh2013}. The EDSA technique has recently been applied to the  fabrication of novel functional nanomaterials, such as photonic colloidal crystals~\cite{Zhou2004}. The EDSA processes can be complicated by the formation of a skin layer at the top of the film, and the formation of irregular structures and vertical non-uniformities~\cite{Reyes2005}. The different mechanisms of film formation during dry coating processes have recently been reviewed~\cite{Felton2013}.
Particle arrangement in the films is mainly controlled by evaporation, particle sedimentation and diffusion processes. To minimize particle sedimentation a controlled temperature gradient has been employed~\cite{Vlasov2001}. The relationship between evaporation and particle diffusion processes can be characterized by the so-called P\'{e}clet number (denoted as $\mathrm{Pe}$) that is the ratio of diffusive time $\tau_D=L_y^2/D$ ($D$ is a diffusion constant of the particle) and the evaporative time $\tau_E=L_y/u$ ($L_y$ is the initial thickness of the film, $u$ is the linear rate of evaporation)~\cite{Routh2013}:
\begin{equation}\label{eq:PeRouth}
\mathrm{Pe}=\tau_D/\tau_E=L_y u/D.
\end{equation}

Where diffusion of the particles is more significant (i.e. at high $D$, $\mathrm{Pe}\ll 1$), a uniform structure of film is expected. In the opposite case, when evaporation dominates over diffusion (i.e. at high $u$, $\mathrm{Pe}\gg 1$), a vertical non-uniformity of the film is expected. For large P\'{e}clet numbers, the theory predicts $\mathrm{Pe}^{1/2}$ scaling in the spatial gradient of the density profile~\cite{Routh2004}.

Particular interest has been paid to simulations of EDSA during the vertical drying of colloidal films. Brownian dynamics have been applied to simulate the vertical drying of binary mixed latex particles dispersed in  water~\cite{Liao2000}. Latex particles were modeled as spheres that interact via the DLVO potential with various surface charge densities. Random spatial distribution and heterocoagulation for small and large differences in the surface potentials between the binary mixed latex particles was observed, respectively. A kinetic MC simulation study of the structure of films filled with colloidal hard spheres during solvent evaporation has also been performed~\cite{Reyes2005}. The study revealed the formation of ordered hexagonal/tetragonal domains and random packing at low and high evaporation rates, respectively. The MC models of latex film formation through evaporation deposition have included the effects of inter-particle and particle-surface interactions~\cite{Reyes2005a}, size polydispersity of the particles and solvent evaporation rate~\cite{Reyes2007}. At low evaporation rates, the formation of films with relatively low porosity and small surface roughness has been observed~\cite{Reyes2007}. Dynamic stratification in drying films of colloidal mixtures of large and small colloidal particles in water has been studied using MC simulation and by experiment~\cite{Fortini2016}. During the drying, the larger particles concentrated at the bottom and the smaller ones concentrated at the top. A segregation mechanism accounting for the presence of a gradient in osmotic pressure was proposed.
A hybrid simulation method based on the combination of classical numerical resolution (using finite differences) and cellular automata methods has been applied to study latex film formation during  vertical drying~\cite{Gromer2015}. Different regimes leading to homogeneous or heterogeneous drying were tested. The distribution of monodisperse latex particles was found to depend on a combination of diffusion, convection, and particle deformation. A linear increase of particle concentration gradient with P\'{e}clet number, $\mathrm{Pe}$, was observed although the effect of increased $\mathrm{Pe}$ was partially hindered in systems with strong repulsive interaction between the particles.

Particle shape anisotropy plays an important role in the drying of a colloidal suspension~\cite{Dugyala2016}. The phenomenon of EDSA and the ordering of anisotropic particles have been experimentally investigated~\cite{Dugyala2013}. The EDSA of hematite nano-ellipsoids with an aspect ratio $2$ and $5$ tunable surface charge has been studied~\cite{Dugyala2015a}. Orientational self-organization of particles has been observed for highly repulsive particle-particle interactions. For weakly charged particles, a disordered structure was obtained. Studies of the drying of colloidal suspensions containing model rod-like silica particles (with aspect ratios ranging from~4 to 15) have revealed parallel orientation of the silica rods in several layers close to the contact line~\cite{Dugyala2015}. Special interest has been paid to the behaviors of thin films obtained by drying colloidal suspensions of carbon nanotubes (CNTs) in experimental investigations (for a review, see~\cite{Hu2010}). The transparent and electrically conductive films of CNTs are of particular  interest in the production of electrodes for super-capacitors, thin film transistors, fuel cells, and battery applications. The presence of EDSA in aqueous CNT suspensions on glass substrates has been reported~\cite{Duggal2006,Sharma2009,Small2006,Li2006}. The EDSA technique has been applied to produce large-area, ordered thin films of CNTs~\cite{Engel2008,Zhang2009,Shastry2013}. In aqueous suspensions of CNTs, isotropic-to-liquid crystalline phase transitions have been observed during water evaporation~\cite{Zhang2009,Shastry2013}. The morphology and coverage of the films could controlled by careful modulation of the CNT and surfactant concentrations.

However, in spite of the great progress which has taken place in experimental investigations of the EDSA phenomenon for anisotropic particles, the theory and computer simulations of such processes have  never previously been discussed in the literature. In earlier studies computer simulation was extensively applied to investigate percolation, jamming and electrical conductivity of films
filled with sticks (continuous problem)~\cite{Balberg1983PRB,Balberg1984PRB,Balberg1987PRA} and by $k$-mers (lattice problem)~\cite{Leroyer1994PRB,Vandewalle2000EPJB,Kondrat2001PRE,Cornette2003epjb,Longone2012PRE,Tarasevich2012PRE,Budinski2016JSM,Lebovka2015PRE,Tarasevich2015JPhCS}. For conducting  anisotropic particles, the particular interest lies in the impact of particle aspect ratio on the electrical conductivity and morphology of the films.

This paper analyzes EDSA for the vertical drying of two-dimensional (2D) colloidal films containing  linear $k$-mers by means of a kinetic MC simulation. The initial state before drying was produced using a model of random sequential adsorption (RSA) on a square lattice with isotropic orientations of the $k$-mers. During the drying, the $k$-mers undergo translation Brownian motion. For completely dried films, the spatial distributions of the $k$-mers and the electrical conductivity in the vertical and horizontal directions were analyzed.

The rest of the paper is constructed as follows. In Section~\ref{sec:methods}, the technical details of the simulations are described, all necessary quantities are defined, and some test results are given. Section~\ref{sec:results} presents our main findings. Section~\ref{sec:conclusion} summarizes the main results.

\section{Computational model\label{sec:methods}}

In the kinetic MC simulation, an RSA model was used to produce an initial homogeneous distribution of rods in the film~\cite{Evans1993RMP}. In the RSA model, overlapping with previously placed $k$-mers is strictly forbidden. The problem was treated on a two-dimensional square lattice of size $L_x\times L_y$ and the rods were represented as linear $k$-mers (i.e., particles occupying $k$ adjacent sites) with a length of $k \in [2, 12]$. The periodic boundary conditions were applied along the horizontal $x$ axis.

The initial concentration was changed within $p_i \in [0, p_j]$, where $p_j$ is the jamming concentration. The jamming concentration $p_j$ corresponds to the state when no one additional $k$-mer can be placed because the presented voids are too small or of inappropriate shape. The values of $p_j$ for different values of $k$ have recently been calculated~\cite{Lebovka2011}. Isotropic orientations of the  $k$-mers (the orientations of $k$-mers along the horizontal $x$ and vertical $y$ directions are equiprobable) was assumed. Vertical drying of the colloidal film was performed along the $y$ axis.

The Brownian diffusion of $k$-mers was simulated using the following kinetic MC procedure (see Appendix~\ref{app:alg}).
We considered fairly dense systems. In this case, rotational diffusion is impeded, especially for large values of $k$. This is the reason why only translational diffusion was taken into consideration in our simulation.
At each step, an arbitrary $k$-mer was chosen and a shift along either the longitudinal ($\parallel$) or the transverse ($\perp$) axis of the $k$-mer on one lattice unit was randomly attempted. The probabilities of translational shifting of the $k$-mers along the different directions $f_\parallel$ and $f_\perp$ were assumed to be proportional to the corresponding coefficients of diffusion, $D_\parallel$ and $D_\perp$.

For a rough approximation, the values of $D_\parallel$ and $D_\perp$ obtained for oblate spheroids were used~\cite{Hoffmann2009}. One can obtain the following equations for the evaluation of $f_\parallel$ and $f_\perp$ (see Appendix~\ref{sec:app} for details):
\begin{equation}\label{EqPr}
     f_\parallel =  \frac{(2/k^2-4)G(k)+2}{(5/k^2-4)G(k)+1},\\
     f_\perp = 1-f_\parallel,
\end{equation}
where $G(k)=\ln[k(1-\sqrt{1-1/k^2})]/\sqrt{1-1/k^2}$.

These equations give slightly different probabilities for a diffusion along the longitudinal or transverse axes, $f_\parallel\geq f_\perp$. For example, for $k=2$ and $k=12$ we have  $f_\parallel\approx 0.534$ and $f_\parallel\approx 0.594$, respectively. One time step of the MC computation, which corresponds to an attempted displacement of all the $k$-mers in the system, was taken as the MC time unit.

\begin{figure}[htbp]
  \centering
  \includegraphics[width=\linewidth,clip=true]{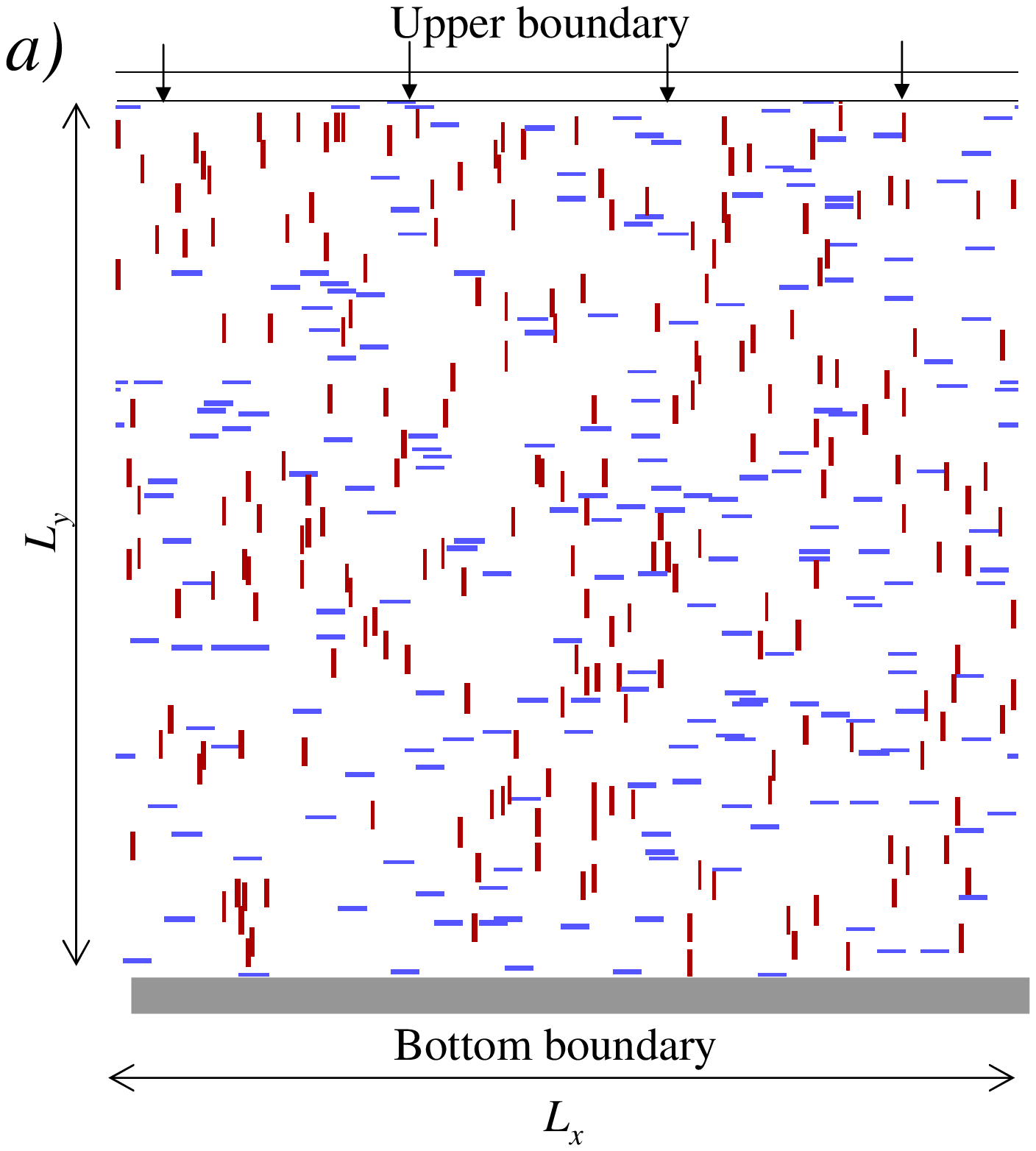}\\
  \includegraphics[width=\linewidth]{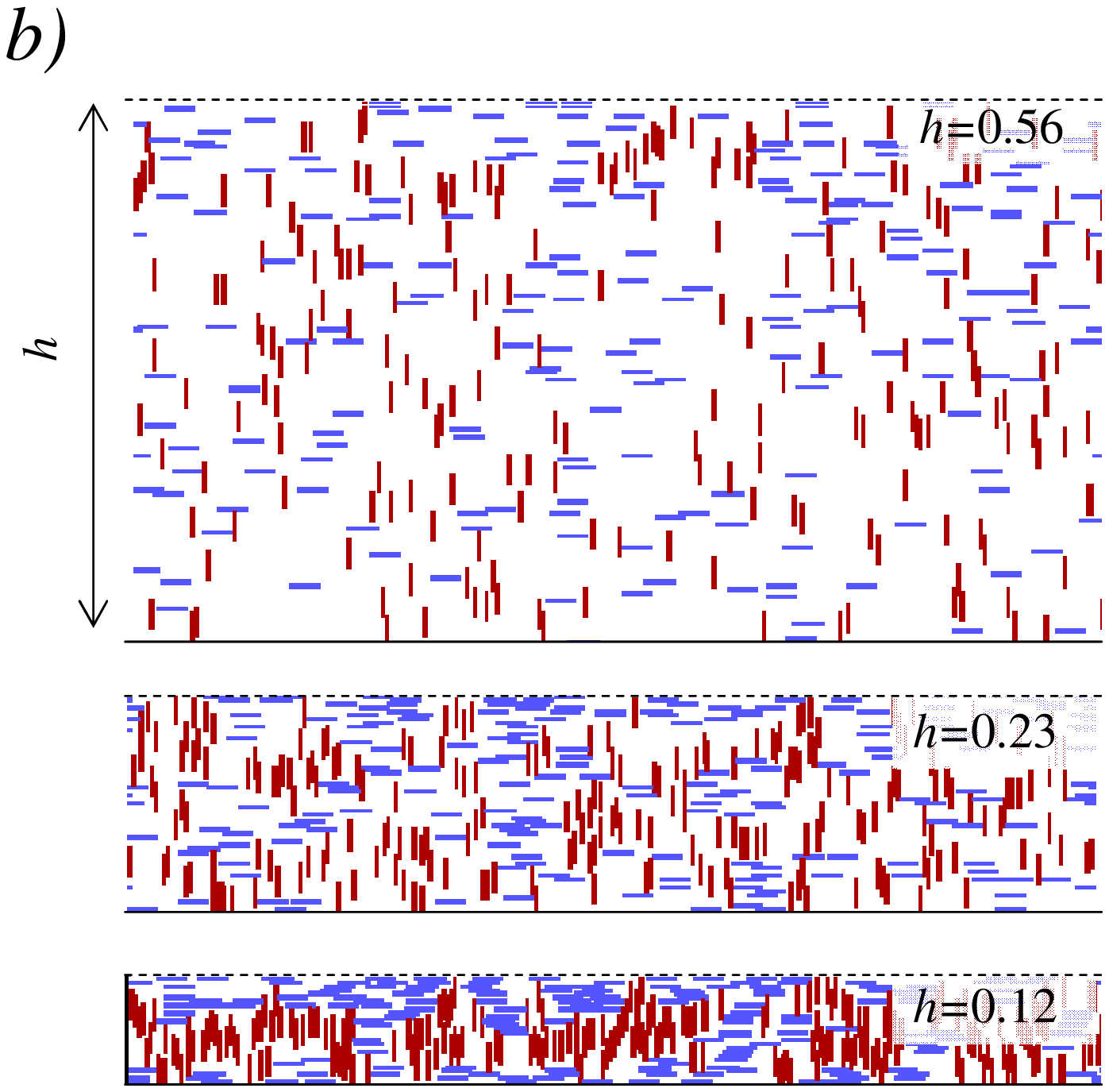}
\caption{(Color online) Examples of $k$-mer configurations at the initial moment before drying (a) and in the process of drying at different values of the relative thickness of the film $h=L_y/L_y^i$:
$h=0.56$ ($t_{MC}\approx 1.7\times 10^{6}$),
$h=0.23$ ($t_{MC}\approx 3\times 10^{6}$)
and $h=0.12$ ($t_{MC}\approx 3.44\times 10^{6}$).
Here, $L_y^i=256$ is the initial size of the film in the vertical direction, $L_x=256$, $k = 8$. The concentration of $k$-mers at the initial moment is $p_i=0.05$, and the evaporation rate is $u=10^{-3}$.
Online: Horizontal $k$-mers are shown in blue, vertical $k$-mers are shown
in red, empty sites are shown in white. Print: gray-scale.\label{fig:Patterns}}
\end{figure}

During drying, the liquid evaporates and the upper interface falls with a linear velocity of $u$ in the vertical  direction (Fig.~\ref{fig:Patterns}a). Short-range repulsion forces between the upper interface and the $k$-mers were assumed and diffusion motions were only allowed inside the film. The drying was stopped when the concentration of $k$-mers in the film was large enough and the diffusion was completely confined to the upper interface of the film. In the absence of $k$-mers the total duration of liquid evaporation is $L_y/u$.

The relationship between the `computational' values $u^c$, $L_y^c$ and the real physical values $u^r$, $L_y^r$ can be established using the equations
\begin{equation}\label{eq:ur}
u^r=u^c d/\tau_B
\end{equation}
and
\begin{equation}\label{eq:Lyr}
L_y^r=L_y^c d,
\end{equation}
where $d$ is the real physical size (diameter) of the monomer (one lattice unit) and
\begin{equation}\label{eq:Lyr1}
\tau_B=3\pi\eta a^3/4kT=d^2/6D
\end{equation}
is the Brownian time, i.e., the time required for particle displacement over a distance $d$.
Here $kT$ is the thermal energy, $\eta$ is the viscosity of evaporating liquid and $D$ is the  diffusion coefficient of the monomer.
The diameter of the monomer (one lattice unit) can be estimated from
\begin{equation}\label{Eq_a}
    d = \sqrt{\dfrac{4kT}{3\pi\eta} \dfrac{u^c}{u^r}}.
\end{equation}

In particular, for the evaporation of water at $T=303$ ($u^r \approx 3.76 \times 10^{-7}$ m/s \cite{Kroeger2007}, $\eta\approx 0.8 \times 10^{-3}$ Pa$\cdot$s)
and a lattice size in the vertical direction of $L_y^c=256$ one can obtain $d \approx 24$ nm and $L_y^r \approx 6$ $\mu$m for $u^c=10^{-4}$, and
$d \approx 80$ nm and $L_y^r \approx 20$ $\mu$m for $u^c=10^{-3}$.
In `computational' units the P\'{e}clet number defined in Eq.~(\ref{eq:PeRouth}) can be represented as
\begin{equation}\label{eq:PeC}
\mathrm{Pe}=L_y^r u^r/D=6L_y^c u^c.
\end{equation}
Hereinafter all `computational' values will be referred to without the upper index $c$.

Figure~\ref{fig:Patterns} presents an example of the $k$-mer configurations at the initial moment before drying (a) and in the process of drying at different values of the relative thickness of the film $h=L_y/L_y^i$, where $L_y^i$ is the initial size of the film in the vertical direction.
The drying resulted in decreasing film thickness, $L_y$, and increasing $k$-mer concentration inside the film, $p$.

\begin{figure}[htbp]
  \centering
  \includegraphics[width=\linewidth]{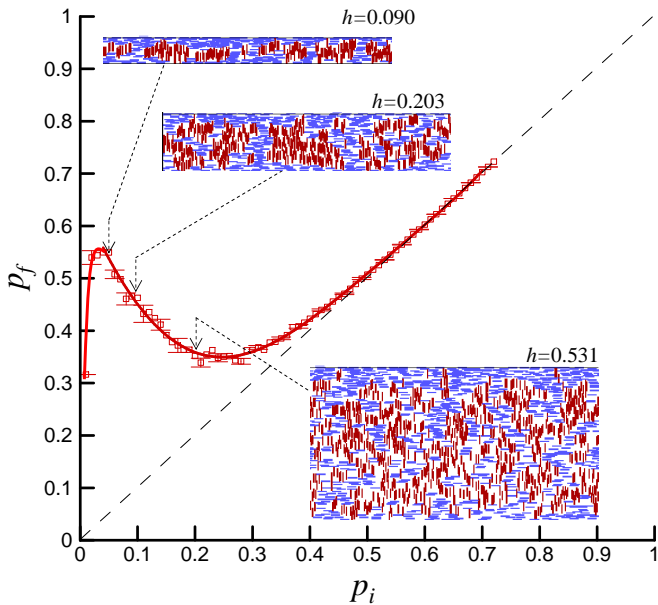}
\caption{(Color online) Example of the final concentration, $p_f$, versus the initial concentration, $p_i$, of the $k$-mers. The patterns of the dried films at different $p_i$ and $h$ are also shown. Here, $L_y^i=256$, $L_x=256$, $h=L_y/L_x$ is the final relative thickness of the dried film,
$k = 8$ is the length of the $k$-mers, and $u=10^{-3}$ is the evaporation rate.
Online: Horizontal $k$-mers are shown in blue, vertical $k$-mers are shown
in red, empty sites are shown in white. Print: gray-scale.\label{fig:Example_pf_pi}}
\end{figure}

The initial concentration $p_i$ affects the final thickness of the dried film $h$ and the final concentration of the $k$-mers within the film $p_f$. Figure~\ref{fig:Example_pf_pi} presents examples of the $p_f$ versus $p_i$ dependence along with the final drying patterns at different values of $p_i$
for $k=8$ and $u=10^{-3}$. For the presented example at small values of $p_i$ ($<0.2$) a significant compaction of the $k$-mers within the film with $p_f>p_i$ is observed. However, at $p_i>0.4$ the compaction is practically absent and ($p_f \approx p_i$).

The profiles of the $k$-mer density, $p(y)$, and the order parameter, $s(y)$, in the vertical direction were evaluated for the films. The order parameter was defined as
\begin{equation}\label{eq:s}
s= \frac{n_y - n_x}{n_x + n_y},
\end{equation}
where $n_x$ and $n_y$  are the number of $k$-mers oriented along the $x$ and $y$ axes, respectively. Here, the case $s=0$ corresponds to an isotropic distribution of the $k$-mers, $s>0$ corresponds to a preferential orientation of $k$-mers along $y$ axis and $s<0$ corresponds to the preferential orientation of $k$-mers along the $x$ axis.

The effects of $k$-mer length and the drying conditions on the electrical conductivity, $\sigma$,  of the films were also studied. For calculation of the electrical conductivity $\sigma_x$ or $\sigma_y$, two conducting buses were applied to the opposite sides of the lattice in the corresponding $y$ or $x$ directions. The electrical conductivity was calculated between these buses. Different electrical conductivities of the bonds between empty sites, $\sigma_m$, filled sites, $\sigma_p$,  and empty and filled sites, $\sigma_{pm}=2\sigma_p \sigma_m / (\sigma_p+\sigma_m)$ were assumed. We put $\sigma_m =1$, and $\sigma_p = 10^6$ in arbitrary units. The Frank-Lobb algorithm was applied to evaluate $\sigma_x$ and $\sigma_y$~\cite{Frank1988PRB}.  Note that $\sigma_x$ or $\sigma_y$ are the  transverse or longitudinal electrical conductivities in a direction perpendicular or parallel to the direction of film drying, respectively. In the calculations, the logarithm of the effective conductivity was averaged over different runs.

\begin{figure}[htbp]
  \centering
  \includegraphics[width=\linewidth]{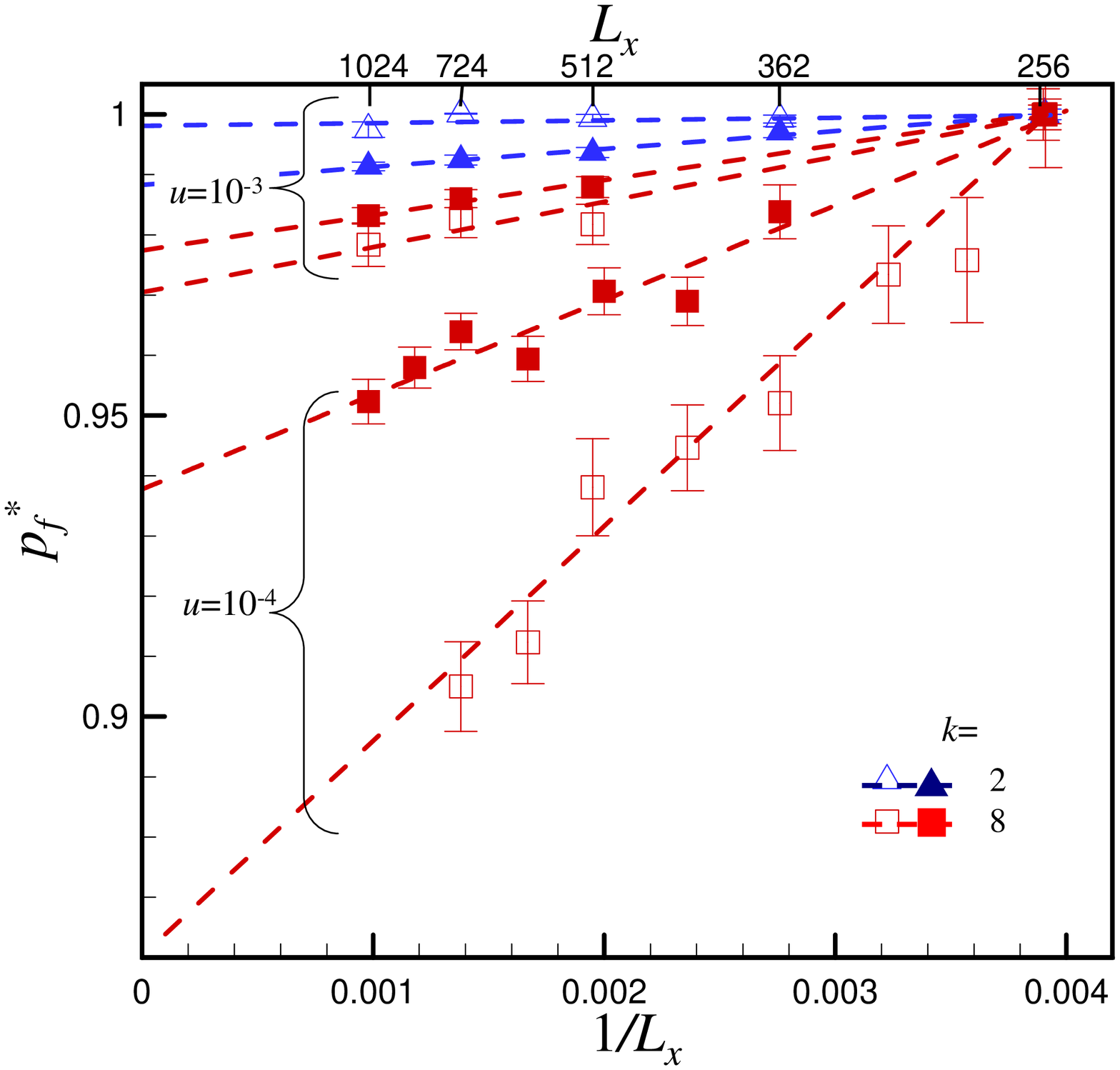}
\caption{(Color online) $p_f^*=p_f(L_x)/p_f(L_x=256)$ versus the $1/L_x$ dependencies at $k=2$ and $k=8$. The empty and filled symbols correspond to
$p_i=0.05$  and $p_i=0.3$, respectively. Here, $L_y^i=256$, and the data were averaged over 100 runs.
\label{fig:Scaling}}
\end{figure}

In the present work, almost all calculation were performed using $L_x=256$, $L_y=256$ and $u=10^{-4}-10^{-3}$. The total MC time required for one drying run, $t_{MC}$ (in MC time units) can be evaluated as
\begin{equation}\label{eq:tMC}
t_{MC}=(1-h)L_y^i/u,
\end{equation}
where $h$ is the final relative thickness of the dried film. This equation gives $t_{MC}\leq 2.56\times (10^{5}-10^{6})$.
A scaling analysis of $p_f$ at a fixed value of $L_y^i=256$ and at different values of $k$ and $L_x$
was performed. Figure~\ref{fig:Scaling} illustrates examples of the scalings of the $p_f$ values.
The difference between the approximated value of $p_f$ in the limit of the infinite system $p_f(L_x \to \infty)$ and $p_f(L_x=256)$ was of the order of several percents for $u=10^{-3}$  and of the order of 10--15 \% for $u=10^{-4}$.
For each given value of $k$, $p_i$ and $u$, the computer experiments were repeated from 100 to 1000 times and the data were averaged.

\section{Results and Discussion\label{sec:results}}
\subsection{Brownian motion driven self-assembly in the absence of drying}
\begin{figure*}
  \centering
  \includegraphics[width=\linewidth]{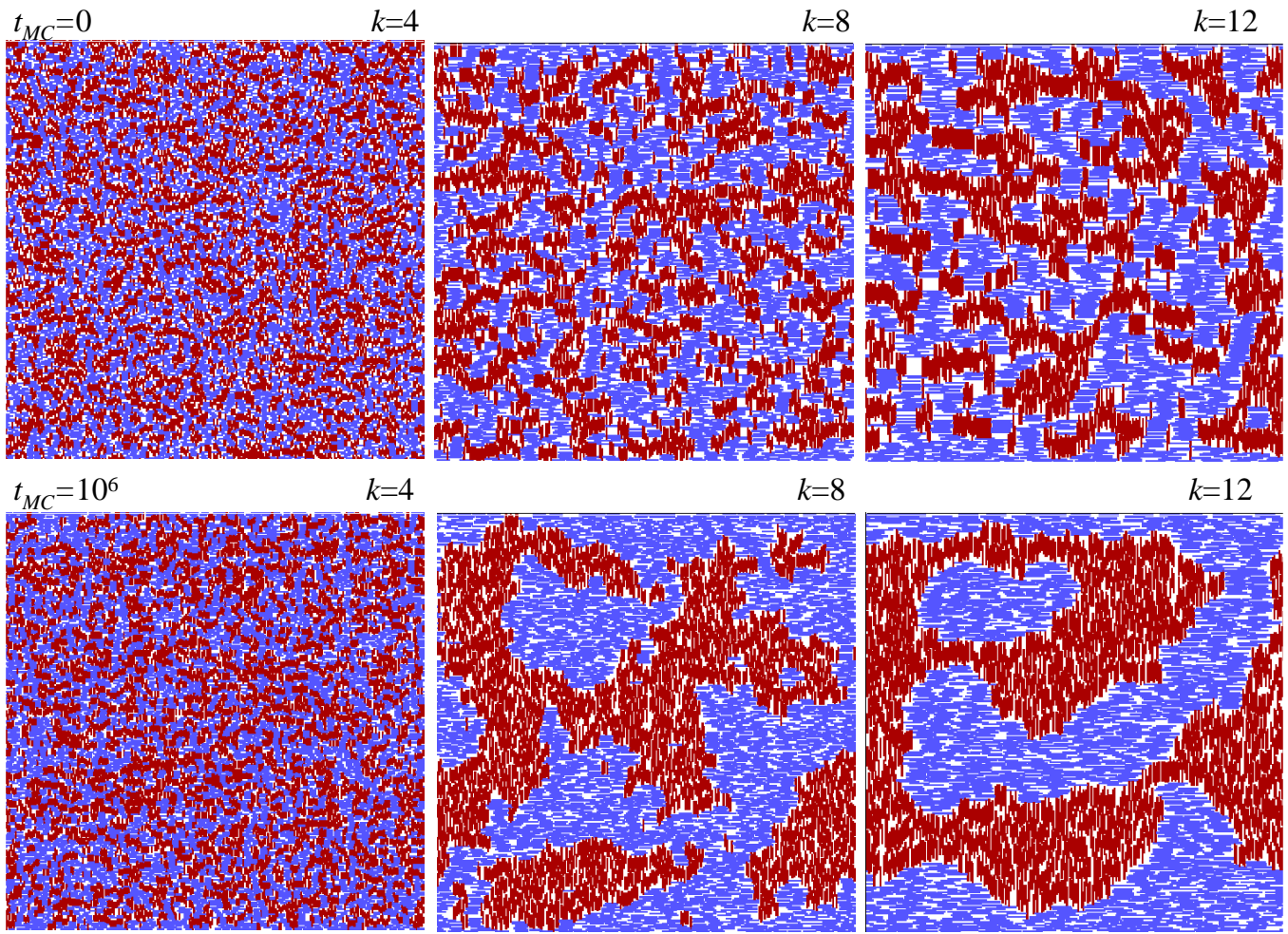}
\caption{(Color online) Examples of $k$-mer patterns at the initial moment ($t_{MC}=0$, upper row) and after $10^6$ MC steps ($t_{MC}=10^6$, bottom row) for $k$-mers of different lengths. Here, $L_y=256$, $L_x=256$, $k$ is the length of the $k$-mers, and evaporation is absent ($u=0$).
Concentration of $k$-mers corresponding to the jamming state: $p\approx 0.8$ ($k=4$),  $p\approx 0.75$ ($k=8$) and $p\approx 0.72$ ($k=12$)~\cite{Lebovka2011}.
Online: Horizontal $k$-mers are shown in blue, vertical $k$-mers are shown in red, empty sites are shown in white. Print: gray-scale.\label{fig:Patternsu0}}
\end{figure*}

The RSA process produces a non-equilibrium state that can be reorganized in the course of the Brownian motion of the $k$-mers.
Figure~\ref{fig:Patternsu0} presents examples of the $k$-mer patterns at the initial moment ($t_{MC}=0$, upper row) and after $10^6$ MC steps ($t_{MC}=10^6$, bottom row) in the absence of evaporation, i.e. at $u=0$. The concentration of $k$-mers was close to the maximum jamming concentration for the given value of $k$~\cite{Lebovka2011}. At the initial moment at $t_{MC}=0$, the deposited $k$-mers tend to align parallel to each other and typical stacks of the horizontally ($x$-stacks) and vertically ($y$-stacks) oriented $k$-mers are observed. These stacks can be represented as squares of size $\approx k \times k$. The jamming state consists of the $x$- and $y$-stacks and the voids between them. These observations are  fully consistent with previously published data~\cite{Manna1991,Vandewalle2000,Lebovka2011}. For randomly oriented $k$-mers (at $s=0$) infinite connectivity (i.e.,  percolation) between the similar $x$- or $y$-stacks has not been observed~\cite{Lebovka2011}. However, a percolation of the similar stacks ($x$ or $y$) can be observed for partially oriented systems.

The differences between the patterns at the initial moment and after $10^6$ MC steps (Fig.~\ref{fig:Patternsu0}) evidenced the presence of a dynamic  spatial reorganization of the system of $k$-mers. In a course of Brownian motion, the horizontal and vertical stacks separate from one another and typical coarsening is observed. With long times, the extent of coarsening becomes much larger than the size of the $k$-mers. Moreover, the horizontal stacks predominantly concentrate near the boundaries whereas the vertical stacks predominantly concentrate in the bulk of the film.

\begin{figure}[htbp]
  \centering
  \includegraphics[width=\linewidth]{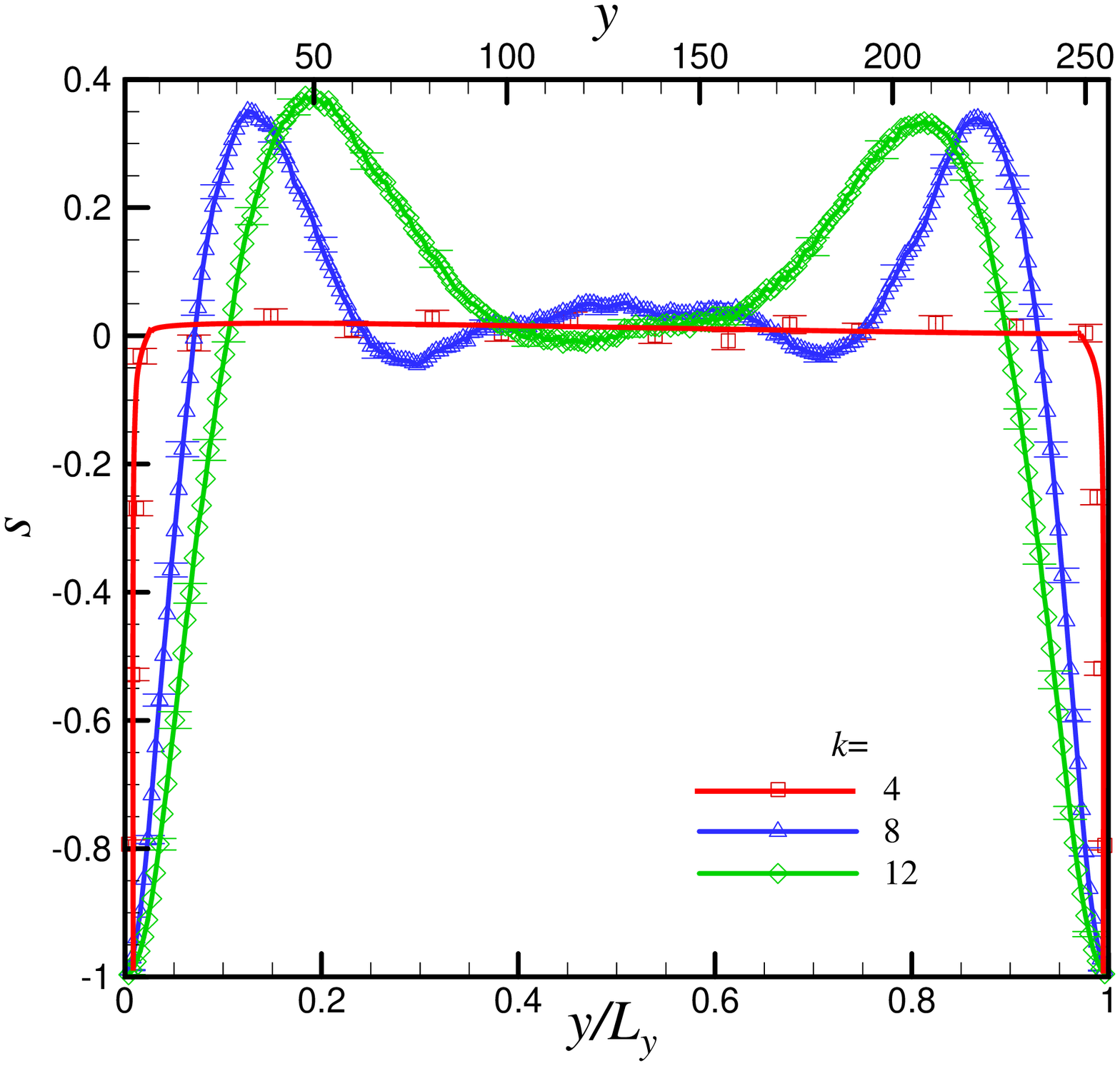}
\caption{(Color online) Examples of the order parameter profiles in the vertical direction, $s(y)$, after $10^6$ MC steps ($t_{MC}=10^6$) for $k$-mers of different lengths. Here, $L_y=256$, $L_x=256$, $k$ is the length of the $k$-mers, and the evaporation is absent ($u=0$).
The concentration of $k$-mers approximately corresponded to that for the jamming state~\cite{Lebovka2011}.\label{fig:Profilesu0}}
\end{figure}

Figure~\ref{fig:Profilesu0} presents examples of the order parameter profiles (Eq.~\ref{eq:s}) in the vertical direction, $s(y)$, after $10^6$ MC steps for $k$-mers of different lengths. The fluctuation of $s(y)$ between negative and positive values increases with increasing length of the $k$-mers.
For example, at $k=8$, the profile in Fig.~\ref{fig:Profilesu0} reveals fluctuations reflecting interleaving of the stacks of $k$-mers oriented in the horizontal and vertical directions. The observed localization of the $x$-stacks near the boundaries evidently reflects the non-periodical boundary conditions in the vertical direction.

For characterization of the degree of segregation , it is useful to introduce the number of inter-contacts of $k$-mers of different orientations, $n$. For example, each site in a vertical (or  horizontal)  $k$-mer can contact $n\leq2$ sites in the horizontal (or vertical) $k$-mers. Complete phase separation corresponds to an absence of contacts between the vertical and horizontal $k$-mers, i.e.,  $n\to 0$.

\begin{figure}[htbp]
  \centering
  \includegraphics[width=\linewidth]{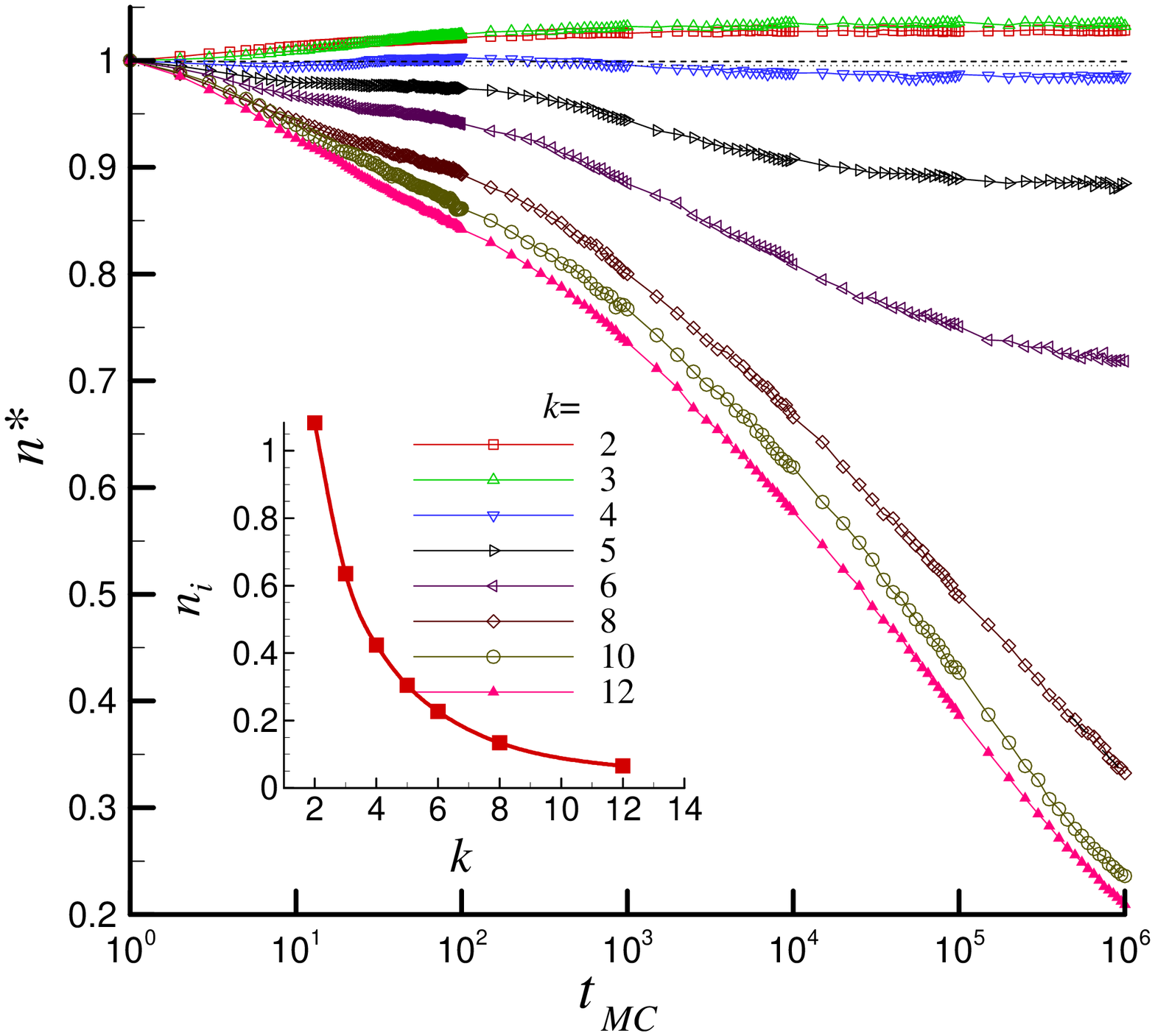}
\caption{(Color online) Relative number of contacts between the vertical and horizontal species, $n^*=n/n_i$ ($n_i$ is the initial number of contacts at $t_{MC}=0$), versus the MC time $t_{MC}$ for different length of $k$-mers. Here, $L_x=L_y=256$. All concentrations of $k$-mer corresponded to the jamming state~\cite{Lebovka2011}. Inset: The initial number of contacts at $t_{MC}=0$ versus the value of $k$.\label{fig:Contacts}}
\end{figure}

Figure~\ref{fig:Contacts} presents the relative number of contacts $n^*=n/n_i$ (here, $n_i$ is the  initial number of contacts at $t_{MC}=0$) versus the MC time $t_{MC}$ for different length of $k$-mers. The initial number of contacts, $n_i$, (see, inset to Fig.~\ref{fig:Contacts}) decreases with increasing length of $k$-mer. The kinetics of the changes of $n$ were different depending on the value of $k$. For $k=2$ and $k=3$, the value of $n$ increases, for $k=3$, the value of $n$ goes through a  maximum at $t_{MC}\approx 100$ and for $k\geq 4$ the value of $n$ decreases  with time $t_{MC}$. It is interesting that for $k \in [2,6]$, the value of $n$ stabilizes at some level for $t_{MC}\approx 10^6$. However, for longer $k$-mers the value of $n$ continues to decrease at $t_{MC}\geq 10^6$. Therefore, for longer $k$-mers, the Brownian motion driven self-assembly requires a longer time for dynamic equilibration. Moreover, the number of contacts between $k$-mers with different orientations in dynamic equilibrium structures significantly decreases with increasing $k$.

It is interesting that different anomalies in the properties of $k$-mer systems have previously been observed in their dependence on the length of the $k$-mers. For example, the jamming concentration decreases monotonically when approaching the asymptotic value of $p_j = 0.66 \pm 0.01$ at large values of $k$, and the percolation threshold $p_c$ is a nonmonotonic function of the length $k$, with a minimum at a particular length of the $k$-mers ($k \approx 13$)~\cite{Kondrat2001PRE,Lebovka2011,Tarasevich2012PRE}. Several problems related to the  self-organization of $k$-mers have been previously discussed~\cite{Ghosh2007,Lopez2010,Loncarevic2010,Kundu2013,Kundu2013a,Matoz-Fernandez2012}. Dynamic Monte Carlo simulations using a deposition-evaporation algorithm for the simulation of dynamic equilibrium of the $k$-mers have been applied~\cite{Ghosh2007}. For long $k$-mers ($k\geq k_m$),  two entropy-driven transitions as a function of density $p$ were revealed: first, from a low-density isotropic phase to a nematic phase with an intermediate density at $p_{in}$, and, second, from the nematic phase to a high-density disordered phase at $p_{nd}$. On a square lattice, $k_m = 7$. A lattice-gas model approach has been applied to study the phase diagram of self-assembled $k$-mers on square lattices~\cite{Lopez2010}. It has been observed that the irreversible RSA process leads to an  intermediate state with purely local orientational order and, in the equilibrium model, the nematic order can be stabilized for sufficiently long $k$-mers~\cite{Matoz-Fernandez2012}. For example, for $k=7$, $p_{in}\approx$~0.729~\cite{Matoz-Fernandez2012} and $p_{nd}\approx 0.917$~\cite{Kundu2013a}.
Thus, the observed Brownian motion driven self-assembly in the absence of drying can reflect an  entropy-driven transition to the high-density disordered phase with separation of the horizontally ($x$-stacks) and vertically ($y$-stacks) oriented $k$-mers.

\subsection{Evaporation driven self-assembly }
To account for the Brownian motion driven self-assembly that was revealed, we can expect a rather complex mechanism of evaporation driven self-assembly for the systems under study. During the drying, the thickness of the film and the concentration of $k$-mers inside it continuously increase with the time.
An initial concentration $p_i$ affects the final concentration of $k$-mers within the film $p_f$ and the final thickness of the dried film $L_y=L_y^ip_i/p_f$.

\begin{figure}[htbp]
  \centering
  \includegraphics[width=\linewidth]{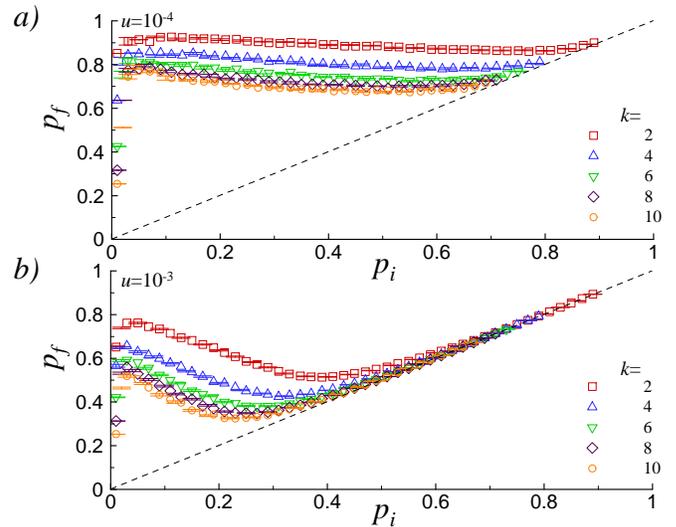}
\caption{(Color online) Final concentration $p_f$ versus initial concentration $p_i$ for $k$-mers of different length and evaporation rate of $u=10^{-4}$ (a) and $u=10^{-3}$ (b). Here, $L_x=256$, $L_y^i=256$, $h=L_y/L_y^i$ is the final relative thickness of the film.
\label{fig:pf vs pi}}
\end{figure}

Figure~\ref{fig:pf vs pi} presents the final concentration $p_f$ versus the initial concentration $p_i$ dependencies  for $k$-mers of different lengths and evaporation rate of $u=10^{-4}$ (a) and $u=10^{-3}$ (b). The final concentration $p_f$ goes through a minimum at a particular concentration of $p_i=p_i^{min}$. The value of $p_i^{min}$ decreases with increasing $k$ and rate of evaporation $u$. The presence of such a minimum reflects the competition between evaporation driven self-assembly and jamming restrictions in a high-density disordered phase.  At a relatively large $p_i$ (at $p_i>0.7$ for $u=10^{-4}$ and $p_i>0.3$ for $u=10^{-3}$), the value of $p_f$ almost coincides with $p_i$, i.e., at high initial density the evaporation driven densification of the film is practically absent.

\begin{figure}[htbp]
  \centering
  \includegraphics[width=\linewidth]{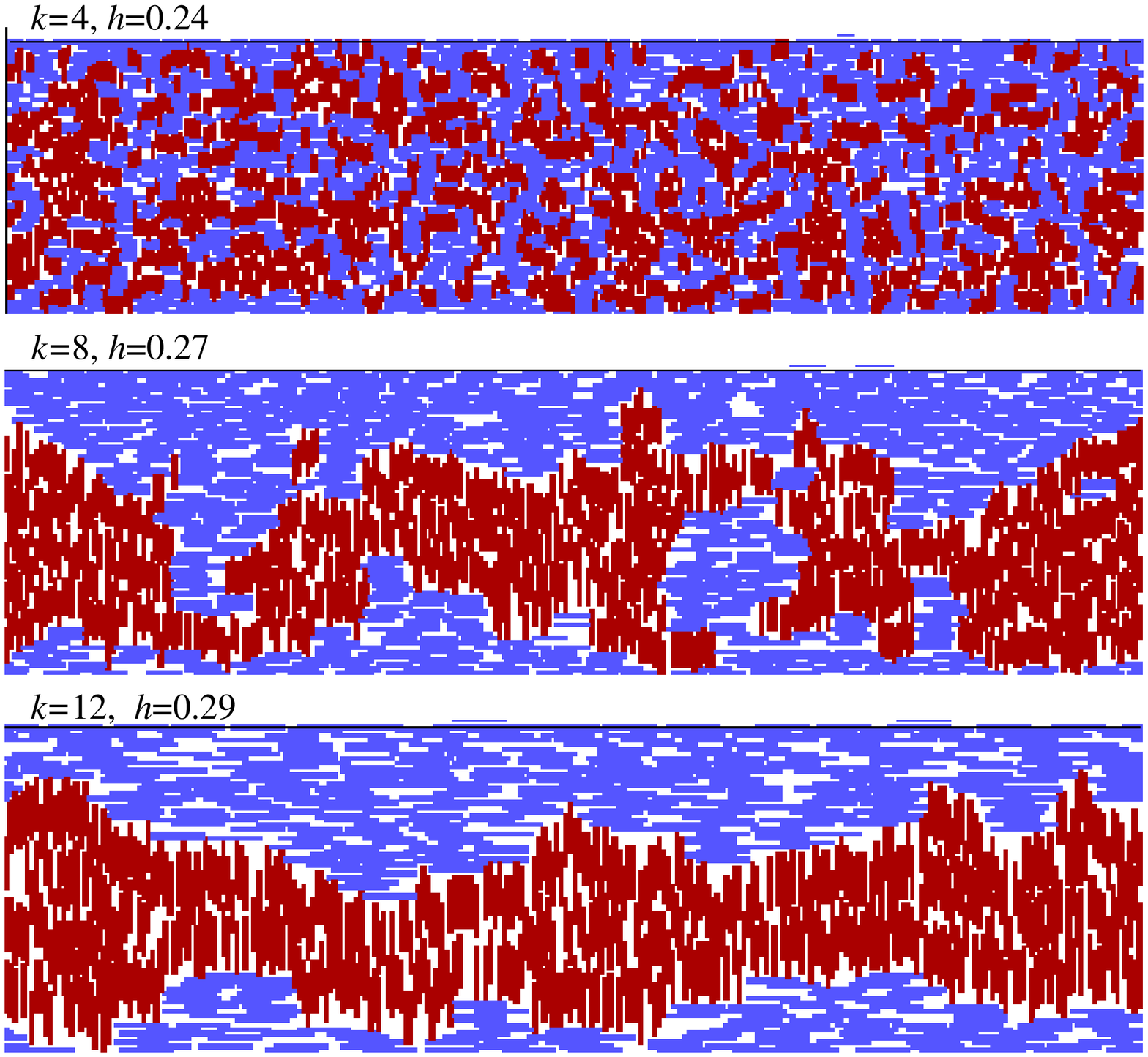}
\caption{(Color online) Examples of $k$-mer patterns at the end of drying for $k$-mers of different lengths. Here, $L_x=256$, $L_y^i=256$, $h=L_y/L_y^i$ is the final relative thickness of the film, $k$ is the length of $k$-mers, and $u=10^{-4}$ is the evaporation rate.
Online: The horizontal $k$-mers are shown in blue, vertical $k$-mers are shown in red, empty sites are shown in white. Print: gray-scale.\label{fig:Patternsu1E4}}
\end{figure}

\begin{figure*}[htbp]
  \centering
  \includegraphics[width=0.45\linewidth]{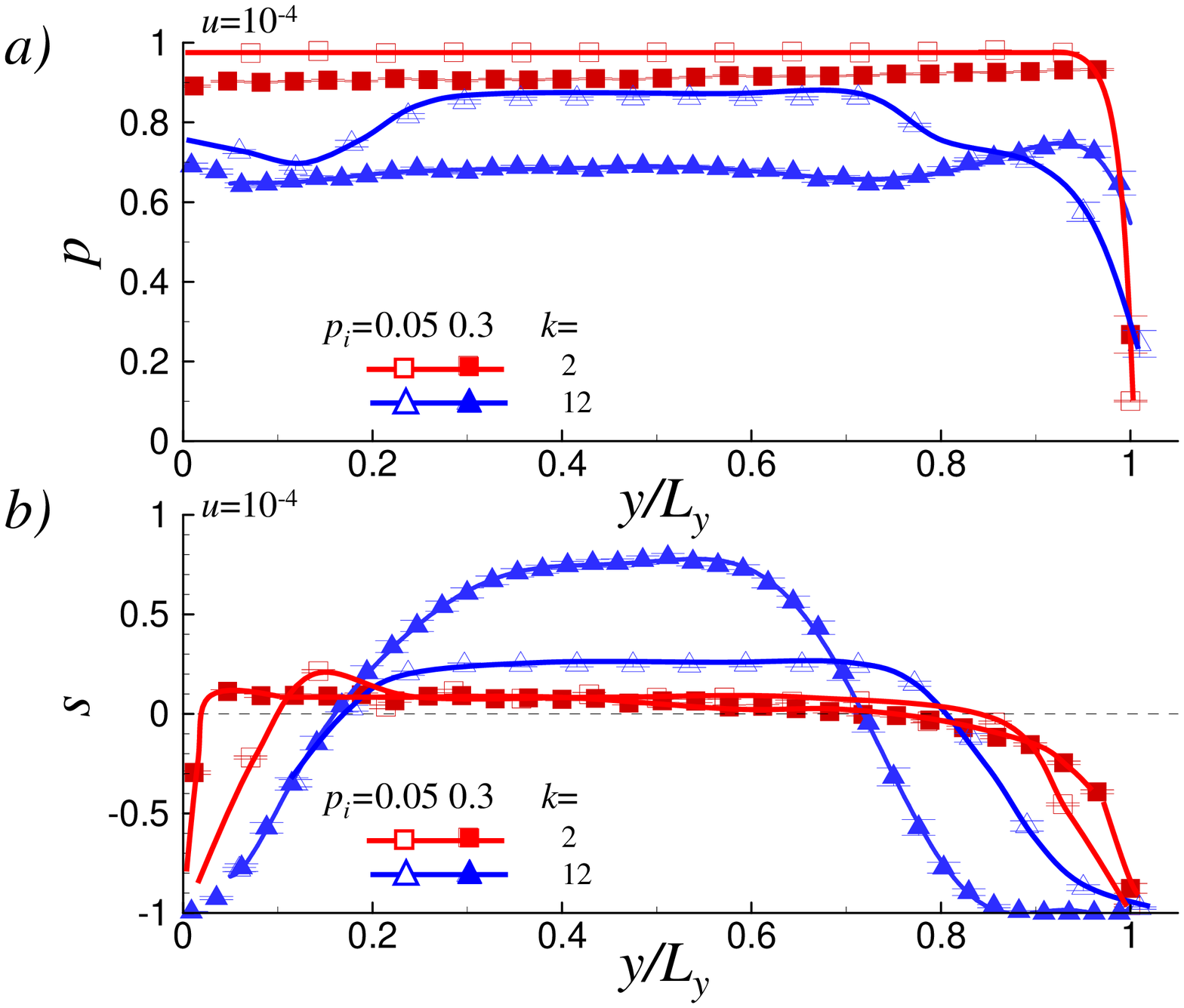}\hfill
  \includegraphics[width=0.45\linewidth]{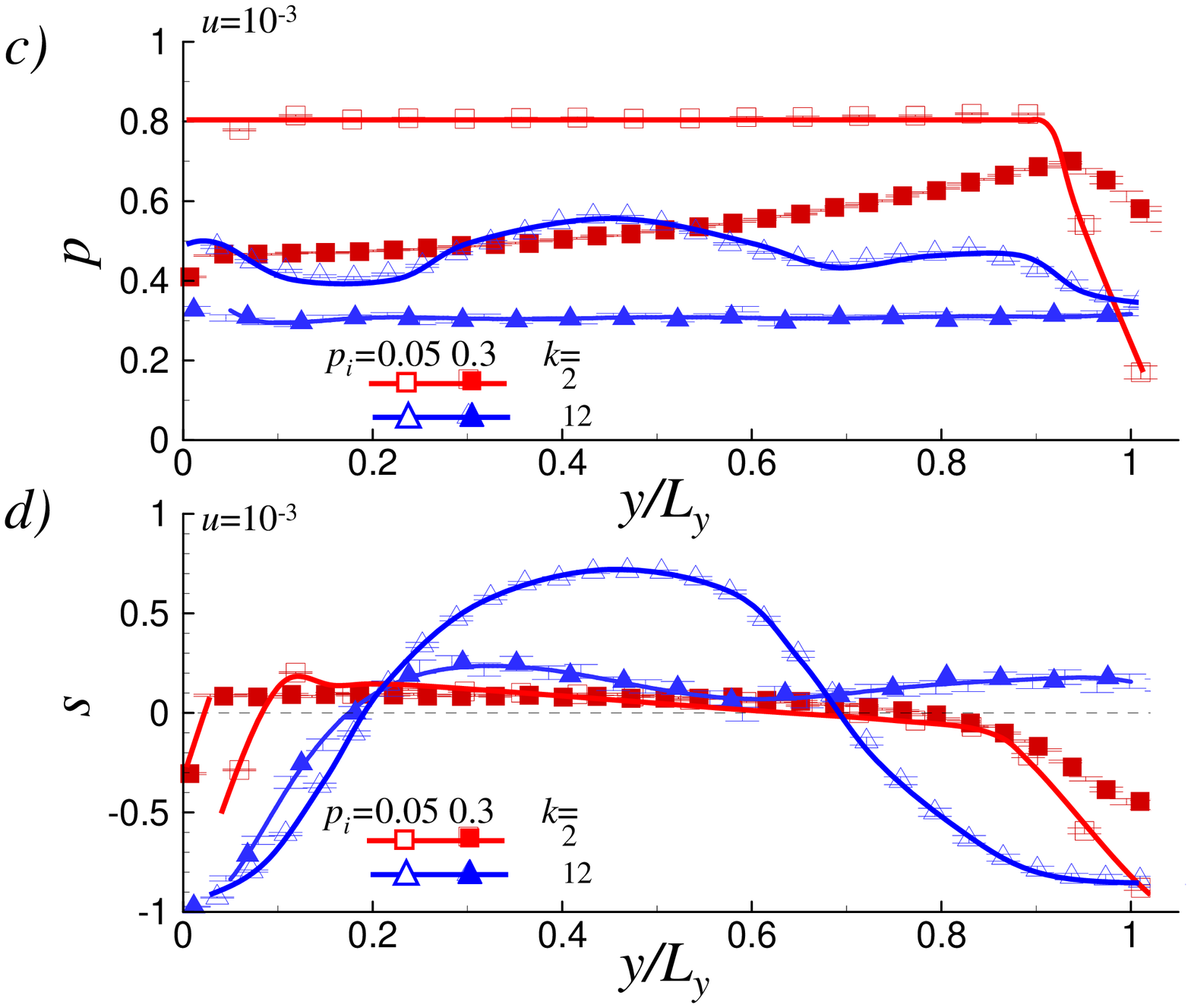}
\caption{(Color online) Profiles of $k$-mer density $p(y)$ (upper) and order parameter $s(y)$ (lower) inside dried films for $k$-mers of different lengths and at evaporation rates of $u=10^{-4}$ (a): $p_i=0.05$, (b): $p_i=0.05$) and $u=10^{-3}$ (d): $p_i=0.05$, (c): $p_i=0.05$). Here, $L_x=256$, $L_y^i=256$.
\label{fig:profiles}}
\end{figure*}

Figure~\ref{fig:Patternsu1E4} presents examples of the $k$-mer patterns at the end of the drying for $k$-mers of different lengths for the particular case of $u=10^{-4}$ and $p_i=0.2$. Here, the stratification of the stacks of $k$-mers oriented in the horizontal and vertical directions is evident. The horizontal $x$-stacks are localized dominantly near the upper and lower boundaries. Moreover,  the density $p(y)$  and order parameter $s(y)$ profiles inside the dried films of $k$-mers (Fig.~\ref{fig:profiles}) evidence the presence of noticeable oscillations in both the $p(y)$  and $s(y)$ functions. An increase in the initial concentration $p_i$ results in a decrease in the density profiles $p(y)$ (Fig.~\ref{fig:profiles}a,c). In some cases, reduction in density near the upper boundary are observed that may reflect the effect of the roughness of the film. For $k=2$ the density oscillations were absent and the profiles $p(y)$ were relatively uniform. In the drying regimes, as used, the P\'{e}clet numbers estimated from Eq.~(\ref{eq:PeC}) are $\mathrm{Pe}\approx 0.154$ at $u=10^{-4}$ and $\mathrm{Pe}\approx 1.54$ at $u=10^{-3}$. For a relatively rapid evaporation rate ($u=10^{-3}$) and a high initial density ($p0.3$) (Fig.~\ref{fig:profiles}c) a spatial gradient in the density profile and the formation of a denser layer near the upper boundary (crust) can reflect a larger P\'{e}clet number. The upper crust consists predominantly of horizontal $x$-stacks with a  negative order parameter (Fig.~\ref{fig:profiles}d). Crust formation has been typically observed during the drying of colloidal suspensions~\cite{Style2011}. For longer $k$-mers the most non-uniform structure of the dried films is observed at small initial concentrations  $p_i$. For example for $k=12$ and $p_i=0.05$ the vertical $y$-stacks with a positive order parameter are predominantly localized in the center of the film (Fig.~\ref{fig:profiles}b,d) where the density of the film is maximal (Fig.~\ref{fig:profiles}a,c).

\begin{figure}[htbp]
  \centering
  \includegraphics[width=\linewidth]{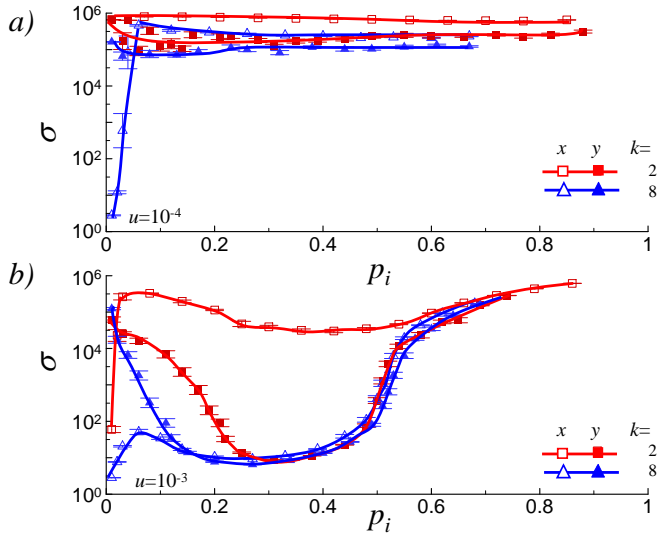}
\caption{(Color online) Electrical conductivity of a dried film $\sigma$ versus initial concentration $p_i$ for $k$-mers of different lengths and evaporation rate of $u=10^{-4}$ (a) and $u=10^{-3}$ (b). The open and filled symbols correspond to the horizontal ($x$) and vertical ($y$) directions. Here, $L_x=256$, $L_y^i=256$.
\label{fig:cond}}
\end{figure}

Figure~\ref{fig:cond} presents the electrical conductivity of a dried film $\sigma$ versus the initial concentration $p_i$ for $k$-mers of different lengths and at evaporation rates of $u=10^{-4}$ (a) and $u=10^{-3}$ (b) in different directions. The open and filled symbols correspond to the horizontal ($x$) and vertical ($y$) directions. It is remarkable that, at small initial concentrations ($p_i<0.05$--$0.1$), the electrical conductivity $\sigma_x$ exceeds the value of $\sigma_y$, whereas at larger values of $p_i$ the opposite behavior is observed. The effects were greatly affected by the length of the $k$-mers. For example, with $k=2$ at a large value of $u$ ($u=10^{-3}$, Fig.~\ref{fig:cond}b) and $p_i$ in the interval between $\approx 0.2$ and $\approx 0.5$  percolation is only observed along the $x$ direction.
Such anisotropy of electrical conductivity may reflect the different connectivity,  stratification, and densification of the stacks of $k$-mers oriented in the horizontal and vertical directions.

\section{Conclusion\label{sec:conclusion}}
A 2D model of the vertical drying of a colloidal film containing rod-like particles ($k$-mers, $k \in [1, 12]$) was studied by means of kinetic Monte Carlo simulation. The initial state before drying was produced by using an algorithm of random sequential adsorption (RSA) with isotropic orientations of the $k$-mers. The simulations were performed at different initial concentrations of the $k$ mers, $p_i$, and values of solvent evaporation velocity, $u$. The chosen drying conditions corresponded to  relatively low P\'{e}clet numbers:  $\mathrm{Pe}\approx 0.154$ at $u=10^{-4}$ and $\mathrm{Pe}\approx 1.54$ at $u=10^{-3}$. During the evaporation, $k$-mers undergo translation Brownian motion, the film thickness continuously decreases, and the density of the $k$-mers inside the film increases. In the completely dried films, the spatial distributions of $k$-mers and the electrical conductivity in the vertical and horizontal directions were analyzed.

A significant stratification of the stacks of $k$-mers oriented along the horizontal and vertical directions was observed. The stratification increases for longer $k$-mers and for larger initial concentrations. Our analysis has shown that the observed evaporation driven self-assembly can include Brownian motion driven self-assembly of the $k$-mers.  Even in the absence of drying, entropy-driven processes of separation of the stacks with different orientations were observed. Meanwhile, the horizontal stacks predominantly concentrated near the boundaries, whereas the vertical stacks predominantly concentrate in the bulk of the film. The development of the drying process influences Brownian motion driven self-assembly by changes in the $k$-mer concentration and the addition of structural non-uniformity in the vertical direction. As a result, the observed evaporation driven self-assembly can be rather different in its dependence on the values of $k$, $p_i$ and $\mathrm{Pe}$.
For example, at $k=2$ and $\mathrm{Pe}\approx 1.54$ the formation of an upper crust of horizontal $x$-stacks with a negative order parameter was observed. For longer $k$-mers ($k=12$), the formation of a more dense layer of vertical $y$-stacks inside the center of film was observed.
A rather intriguing behavior of the electrical conductivity of dried films was observed. Films with significant anisotropy of electrical conductivity in the horizontal and vertical directions can be obtained. The anisotropy can be finely regulated by changes in the values of $p_i$, $k$ and $u$. Such anisotropy of electrical conductivity evidently reflects the different connectivity,  stratification, and densification of the stacks of $k$-mers oriented in the horizontal and vertical directions.

\section*{Acknowledgements}
The authors would like to thank Andrei S. Burmistrov for his technical assistance.
We also acknowledge the funding from the National Academy of Sciences of Ukraine, Project No.~43/16-H (NL, NV, and VG) and the Ministry of Education and Science of the Russian Federation, Project No.~643 (Yu.T).

\appendix
\section{Algorithm\label{app:alg}}

\begin{algorithmic}[1]
\REPEAT
\STATE{}
\COMMENT{One Monte Carlo step}
\FOR{$i = 1$ to number of $k$-mers to be shifted}
\STATE{Randomly select a $k$-mer}
\STATE{Randomly select a shift direction}
\IF{The shift direction coincides with the orientation of the $k$-mer}
probability of shift $\gets 1$
\ELSE{ probability of shift $\gets f$}
\ENDIF
\STATE{Try to shift the $k$-mer in the chosen direction by one lattice site with the given probability}
\ENDFOR
\STATE{Move upper boundary of the film to a new position according evaporation rate}
\UNTIL{Drying is finished}
\end{algorithmic}

\section{Evaluation of the probabilities of diffusive motion along different directions $f_\parallel$ and $f_\perp$ \label{sec:app}}

The probabilities of translational shifting of $k$-mers along the different directions $f_\parallel$ and $f_\perp$ were assumed to be proportional to the corresponding coefficients of diffusion, $D_\parallel$ and $D_\perp$.
For prolate ellipsoids of revolution (spheroids) with two semiaxes of equal length $b$ and a long semiaxis of length $a$, the translational diffusion coefficients parallel ($D_\parallel$) and perpendicular ($D_\perp$) to the main symmetry axis are given by
\begin{equation}\label{eq:dpar}
D_\parallel = \frac{k_B T}{8 \pi \eta}  \frac{ \left( 2 – r^2 \right) G( r ) -1}{ 1 - r^2} ,
\end{equation}
\begin{equation}\label{eq:derp}
D_\perp = \frac{k_B T}{16 \pi \eta}  \frac{ \left( 2 – 3r^2 \right) G( r ) + 1}{ 1 - r^2} ,
\end{equation}
where $r = b/a<1$ is the aspect ratio, $k_B$ is the Boltzmann constant, $T$ is the temperature, $\eta$ is the viscosity, and $G(r) = \ln \left(\left( 1 +\sqrt{ 1 - r^2}\right) \right) / \sqrt{ 1 - r^2}$ \cite{Hoffmann2009}.
In the limiting case of spherical particles ($r \to 1$, $b =a$), these diffusion coefficients are equal to the well-known Einstein-Stokes formula
\begin{equation}\label{eq:ESF}
D_\perp = D_\parallel= \frac{k_B T}{6 \pi \eta b}.
\end{equation}

In our model, a $k$-mer can be considered as a prolate ellipsoid of revolution with two equal semiaxes of length $b=1/2$ and a long semiaxis with a length of $a = k/2$. In this case, the orientation of the $k$-mer coincides with the orientation of the long semiaxis of the ellipsoid and the aspect ratio $r = 1/k$.
We suppose that the probabilities $f_\parallel$, $f_\perp$ of $k$-mer shift along ($\parallel$) and perpendicular ($\perp$) of the $k$-mer orientation are proportional to the corresponding diffusion coefficients of the prolate spheroid $f_\parallel = c D_\parallel$, $f_\perp = c D_\perp$, where $c$ is a normalization factor.
The relationship
\begin{equation}\label{EqPrA}
     f_\parallel = \frac{D_\parallel}{ D_\parallel + D_\perp}
\end{equation}
yields \eqref{EqPr}.

\bibliography{Drying}

\end{document}